\documentclass{article}
\usepackage{spconf,amsmath,epsfig,amsfonts,amssymb,psfig,graphics,psfrag,theorem,calc,subfigure,url,bm,cite }
\usepackage{amsmath,multirow,epsfig,amsfonts,amssymb,psfig,graphics,psfrag,theorem,calc,subfigure,url,bm,cite}
\usepackage{graphicx,color}  
\usepackage{stfloats}
\usepackage{algorithmic,algorithm}
\usepackage{shortcuts_OPT}

\usepackage{calc,amsfonts,amssymb,amsmath,bm,url,color,theorem,graphicx,cite}
\usepackage{psfrag,subfigure,float}
\usepackage{soul}
\usepackage{bbm}
\usepackage{enumerate}

\newlength{\picwidth}
    \makeatletter
    \def\multilimits@{\bgroup
  \Let@
  \restore@math@cr
  \default@tag
 \baselineskip\fontdimen10 \scriptfont\tw@
 \advance\baselineskip\fontdimen12 \scriptfont\tw@
 \lineskip\thr@@\fontdimen8 \scriptfont\thr@@
 \lineskiplimit\lineskip
 \vbox\bgroup\ialign\bgroup\hfil$\m@th\scriptstyle{##}$\hfil\crcr}
    \def\Sb{_\multilimits@}
    \def\endSb{\crcr\egroup\egroup\egroup}
\makeatother

{\begin{list}{}%
    {\setlength{\rightmargin}{0pc}%
    \setlength{\leftmargin}{1pc} }} %
{\end{list}}

\newlength{\twidth}

\newtheorem{Lemma}{Lemma} 

\newtheorem{Theorem}{Theorem} 

\theorembodyfont{\rmfamily}

\newtheorem{Question}{Question}

{\begin{list}{}%
    {\setlength{\rightmargin}{1pc}%
    \setlength{\labelsep}{0pc}
    \setlength{\leftmargin}{1pc} }} %
{\end{list}}

{\begin{list}{}%
    {\setlength{\rightmargin}{0pc}%
    \setlength{\leftmargin}{0pc} }} %
{\end{list}}

{\begin{list}{}{
    \settowidth{\labelwidth}{\mbox{\textnormal{#1}}}%
    \setlength{\leftmargin}{\labelwidth+\labelsep}}}%
{\end{list}}

\newsavebox{\savepar}

\ninept

\newcommand\bx{\ensuremath{{\bm x}}}
\newcommand\by{\ensuremath{{\bm y}}}
\newcommand\bG{\ensuremath{{\bm G}}}

\newcommand\bH{\ensuremath{{\bm H}}}

\newcommand\be{\ensuremath{{\bm e}}}
\newcommand\bz{\ensuremath{{\bm z}}}
\newcommand\bM{\ensuremath{{\bm M}}}

\newcommand\bX{\ensuremath{{\bm X}}}

\newcommand\ba{\ensuremath{{\bm a}}}

\newcommand\bA{\ensuremath{{\bm A}}}

\newcommand\bb{\ensuremath{{\bm b}}}

\newcommand\bB{\ensuremath{{\bm B}}}

\newcommand\balp{\ensuremath{{\bm \alpha}}}

\newcommand\bF{\ensuremath{{\bm F}}}

\newcommand\bd{\ensuremath{{\bm d}}}

\newcommand\bv{\ensuremath{{\bm v}}}

\newcommand\btheta{\ensuremath{{\bm \theta}}}

\newcommand\bY{\ensuremath{{\bm Y}}}

\newcommand\bU{\ensuremath{{\bm U}}}
\newcommand\bs{\ensuremath{{\bm s}}}
\newcommand\bS{\ensuremath{{\bm S}}}

\newcommand{\Rbb}{\mathbb{R}}

\newcommand{\setA}{\mathcal{A}}

\newcommand{\setX}{\mathcal{X}}

\newcommand{\setU}{\mathcal{U}}

\newcommand\br{\ensuremath{{\bm r}}}

\newcommand{\bzero}{{\bm 0}}
\newcommand{\bone}{{\bm 1}}
\newcommand{\bI}{{\bm I}}

\newcommand\indfn[1]{{{\mathbbm 1}_{#1}}}

\newcommand\sspan{\ensuremath{{\rm span}}}
\newcommand\aff{\ensuremath{{\rm aff}}}
\newcommand\conv{\ensuremath{{\rm conv}}}

\newcommand\svol{\ensuremath{{\rm vol}}}

\definecolor{orange}{RGB}{255,107,0}

\title{On Hyperspectral Unmixing}

\name{Wing-Kin Ma \vspace{-0.3cm}
\thanks{The work was supported by a General Research Fund (GRF) of Hong Kong Research Grant Council (RGC) under Project ID CUHK 14205717.}}
\address{Department of Electronic Engineering, The Chinese University of Hong Kong, Hong Kong SAR of China}

\begin{document}

\maketitle \sloppy

\begin{abstract}
In this article the author reviews Jose Bioucas-Dias’ key contributions 
to hyperspectral unmixing (HU), in memory of him as an influential 
scholar and for his many beautiful ideas introduced to the hyperspectral 
community. Our story will start with vertex component analysis 
(VCA)---one of the most celebrated HU algorithms, with more 
than 2,000 Google Scholar citations. VCA was pioneering, invented at a 
time when HU research just began to emerge, and it shows sharp insights 
on a then less-understood subject. Then we will turn to SISAL, another 
widely-used algorithm.
SISAL is not only a highly successful algorithm, it is also a 
demonstration of its inventor’s ingenuity on applied optimization and on 
smart formulation for practical noisy cases. Our tour will end with 
dependent component analysis (DECA), perhaps a less well-known 
contribution. DECA adopts a statistical inference framework, and the 
author’s latest research indicates that such framework has great 
potential for further development, e.g., there are hidden connections 
between SISAL and DECA. The development of DECA shows foresight years 
ahead, in that regard.
\end{abstract}

\section{Introduction}


It is well-known within the hyperspectral community that hyperspectral unmixing (HU) is a promising topic, providing a strikingly rich variety of methods for blindly identifying materials' spectral responses from a hyperspectral image.
Since HU is a well-established topic, the author will not elaborate on, or restate, the numerous developments of HU; the reader can easily find overview articles such as \cite{Jose12,Ma2014HU} and closely-related articles such as 
\cite{fu2019nonnegative,gillis2021nmf}.
The author wants to give a short tour on Jos\'{e} Bioucas-Dias’ original contributions to HU, in memory of him as the undisputedly greatest researcher of our time in hyperspectral signal and image processing.
The author's background is on signal processing, with an emphasis on fundamental aspects.
It is inevitable that he will use his lens to view Bioucas-Dias’ inventions and the insights thereof,
and the reader should note that his view represents only one of the perspectives to appreciate Bioucas-Dias’ works.
In fact, the author and Bioucas-Dias did not necessarily share the same view on every aspect despite the fact that they are good friends and had many discussions for about a decade.

This paper will cover vertex component analysis (VCA) \cite{Nascimento2005}, simplex identification via split augmented Lagrangian (SISAL) \cite{Dias2009} and dependent component analysis (DECA) \cite{nascimento2012hyperspectral}.
VCA is very widely-used and should be the most cited work in the history of HU.
SISAL is arguably the most popularly-used algorithm among the various simplex volume minimization algorithms in HU.
DECA is, by comparison, not as well-known, but it shows significant insights and great potential as the author will explain.
The author will not cover the model-order estimator HySime and sparse unmixing, which are also  Bioucas-Dias' key contributions.

The author assumes the prerequisite that the reader has knowledge about the signal processing basics of HU (see, e.g., \cite{Ma2014HU,fu2019nonnegative,gillis2021nmf}).
Or, for a reader not from remote sensing, he or she should be equipped with relevant concepts in signal processing or machine learning, e.g., blind source separation, 
or non-negative matrix factorization.

\section{VCA}
\label{sect:vca}

We begin by hypothesizing that the hyperspectral image we capture obeys a noiseless linear mixture model
\beq \label{eq:model}
\by_t = \textstyle \sum_{i=1}^N \ba_i s_{i,t} = \bA \bs_t, \quad t=1,\ldots,T,
\eeq 
where 
$\by_t \in \Rbb^M$ collects the reflectances of the image over $M$ spectral bands and at a pixel indexed by $t$;
each $\ba_i \in \Rbb^M$ is the spectral response of a distinct endmember;
$\bA = [~ \ba_1, \ldots, \ba_N ~]$;
$\bs_t = [~ s_{1,t}, \ldots, s_{N,t} ~]^\top$ describes the abundances of the different endmembers at pixel $t$;
$N$ is the number of endmembers;
$T$ is the number of pixels.
It is typical to assume, and we will assume, that 
i) every $\bs_t$ lies in the unit simplex $\setU := \{ \bs \in \Rbb^N \mid \bs \geq \bzero, 
\bs^\top \bone = 1
 \}$;
ii) $\bA$ has full column rank;
iii) $\bS = [~ \bs_1,\ldots, \bs_T ~]$ has full row rank.
The problem of HU is to identify $\bA$ from the image $\bY = [~ \by_1, \ldots, \by_T ~]$.

VCA uses the pure-pixel assumption---i.e., for each endmember, there exists a pixel that is contributed purely by that endmember. 
Mathematically, we say that the pure-pixel assumption holds if, for each $i$, there exists a $t_i \in \{1,\ldots, T \}$ such that $\by_{t_i} = \ba_i$.
Under the pure-pixel assumption, the problem of identifying $\bA$ can be done by finding $t_1,\ldots,t_N$.
VCA has its insight reminiscent of Boardman's pure-pixel index (PPI) \cite{PPI}, which employs
\beq \label{eq:rand_proj}
\hat{t} \in \arg \max_{t=1,\ldots,T} | \langle \by_t, \br \rangle |,
\eeq 
to identify pure pixels.
Here, $\langle \cdot, \cdot \rangle$ denotes the inner product;
$\br$ is a randomly drawn direction.
Eq.~\eqref{eq:rand_proj} randomly projects the $\by_t$'s into a line and then finds an extreme point there.
It is easy to show that $\hat{t}$ is one of the pure-pixel indices $t_i$'s, with a high probability.
The issue with \eqref{eq:rand_proj} is that 
we may need to re-run \eqref{eq:rand_proj} many many times to obtain $t_1,\ldots,t_N$, {\em all} endmembers' pure-pixel indices.

VCA fixes the above issue by orthogonal projection.
Suppose we already identified a number of $k$ pure-pixel indices, each corresponding to a distinct endmember. 
Without loss of generality (w.l.o.g.), let $t_1,\ldots,t_k$ be the identified pure-pixel indices.
Since $\ba_1,\ldots, \ba_k$ is known (note $\by_{t_i}= \ba_i$),
consider a randomly drawn direction $\br_k$ such that $\br_k$ is orthogonal to $\ba_1,\ldots,\ba_k$.
Since
\beq
| \langle \by_t, \br_k \rangle | = | {\textstyle \sum_{i=k+1}^N \langle \ba_i, \br_k \rangle s_{i,t} } | \leq \max_{i=k+1,\ldots,N} | \langle \ba_i, \br_k \rangle |,
\eeq 
where equality holds if $\by_t = \ba_l$, $l \in \arg \max_{i=k+1,\ldots,N} | \langle \ba_i, \br_k \rangle |$,
it makes sense to consider 
\beq \label{eq:vca}
\hat{t}_{k+1} \in \arg \max_{t=1,\ldots,T} | \langle \by_t, \br_k \rangle |
\eeq 
to identify a new pure-pixel index.
One may show that $\hat{t}_{k+1}$ is a pure-pixel index associated with a new endmember $\ba_i$, $i \in \{ k+1,\ldots,N \}$, with a high probability.
The idea of VCA is to successively run \eqref{eq:vca}, from $k=1$ to $k=N$.
Compared with PPI, VCA requires calling 
the projections 
$N$ times only.

The merit of VCA is that it is computationally very efficient.
While later developments lead to similar algorithms, with some having better fundamental explanations and/or provable identifiability guarantees (see, e.g.,  
\cite{fu2019nonnegative,gillis2021nmf} 
and the references therein) ,
VCA was invented at a time when the theory and methods for pure-pixel search were much less well-understood than what we know today.

\section{SISAL}

SISAL is an algorithmic realization of simplex volume minimization (SVMin), an idea first conceived by Craig in HU in 1994 \cite{Craig1994}.
SVMin was an intuition that said that  the endmembers can be identified by finding a minimum-volume simplex that encloses all the hyperspectral pixels.
This intuition was empirically found to be valid in later studies, even for heavily mixed pixel (and no-pure-pixel) instances;
and it is recently confirmed to be true in theory, under some assumptions \cite{lin2015identifiability,Fu2015}.
It is now commonly accepted that SVMin can be mathematically described by an optimization problem
\beq \label{eq:svmin}
\begin{aligned}
	\min_{\bA} & ~ \svol(\bA): = (\det(\bar{\bA}^\top \bar{\bA}))^{1/2}/(N-1)! \\
	{\rm s.t.} & ~ \by_t \in \conv(\bA):= \{ \by= \bA \bs \mid \bs \in \setU \}, ~ t=1,\ldots,T
\end{aligned}
\eeq 
where $\bar{\bA} := [~ \ba_1 - \ba_N, \ldots, \ba_{N-1} - \ba_N ~]$,
and the solution to problem~\eqref{eq:svmin} serves as the endmembers estimate.
Here, $\conv(\bA)$ denotes the convex hull of the set of points $\ba_1,\ldots, \ba_N$,
which is a simplex under the assumption of affinely independent $\ba_1,\ldots, \ba_N$;
$\svol(\bA)$ is the volume of the simplex $\conv(\bA)$.
There was an issue back in the 2000's---from Craig's 1994 paper it was not clear how the SVMin problem was solved, precisely.

\subsection{Formulation}

Bioucas-Dias was the among the first who seriously studied the realization of the SVMin problem \eqref{eq:svmin}.\footnote{The author's team happened to begin to investigate SVMin around the same time as Bioucas-Dias' team,
	and the co-occurrence of the two independent research led the two sides to know each other in 2009 WHISPERS.
	The author is grateful to Tsung-Han Chan, the key member of the author's team back then. He made bold attempts to study HU---which was then unknown to the team---and triggered the team's interest in SVMin.}
Let us first describe the formulation \cite{Li2008}.
Consider the following:
i) $M= N$ such that $\bA$ is square;
ii) replace $\det(\bar{\bA}^\top \bar{\bA})$ in \eqref{eq:svmin} by 
$\det({\bA}^\top {\bA})= | \det(\bA) |^2$.
Then we can formulate the SVMin problem as
\beq \label{eq:svmin_jose}
	\min_{\bA, \bS}  ~ |\det(\bA)| \quad 
	{\rm s.t.}  ~ \bY = \bA \bS, ~ \bS \geq \bzero, ~  \bS^\top \bone  = \bone.
\eeq 
By a change of variable $\bB = \bA^{-1}$, we can transform \eqref{eq:svmin_jose} as
\beq \label{eq:svmin_jose2}
	\max_{\bB}  ~  |\det(\bB)|  \quad 
	{\rm s.t.}  ~ \bB \bY \geq \bzero, ~ \bB^\top \bone =  (\bY^\top)^\dag \bone,
\eeq 
where the superscript $\dag$ denotes the pseudo-inverse.
The equivalence of problems \eqref{eq:svmin_jose} and \eqref{eq:svmin_jose2} is shown in Section~6.1.\footnote{Bioucas-Dias wrote down the problem transformation \eqref{eq:svmin_jose2} very concisely (see \cite{Li2008}).
The author has been long wondering if a mathematically precise proof on the equivalence of the transformation can be provided.
}
Problem \eqref{eq:svmin_jose2} is easier to handle than problem \eqref{eq:svmin_jose} because the former's constraints are convex.
However the constraint $\bB \bY \geq \bzero$ is a number of non-separable inequality constraints,
and their presence poses limitations to the development of computationally efficient schemes for \eqref{eq:svmin_jose2}.
In that regard we should mention that $T$, the number of pixels, is large, and we are dealing with an optimization problem that has numerous non-separable inequality constraints.
As a compromise,  Bioucas-Dias turned to a soft-constrained variant of \eqref{eq:svmin_jose2}
\beq \label{eq:svmin_jose3}
\min_{\bB}  ~  f(\bB) + \lambda \, h(\bB \bY)  \quad 
{\rm s.t.}  ~ \bB^\top \bone =  (\bY^\top)^\dag \bone,
\eeq
where $\lambda > 0$ is given;
\[
f(\bB):= -\log(|\det(\bB)|); \quad h(\bX):= \textstyle \sum_{i,j} \max\{ -x_{ij}, 0 \}.
\]
Here, $h(\bX)$ is an element-wise hinge function, serving as a penalizer to discourage $\bX$ from having negative elements;
the incorporation of $\log$ on $|\det(\bB)|$, which is w.l.o.g., is to make the problem numerically better to tackle---intuitively.

\subsection{Optimization}

We now turn to the optimization.
In  SISAL, successive convex optimization is adopted to tackle problem \eqref{eq:svmin_jose3}, specifically,
\beq
\label{eq:sca_sisal}
\bB^{k+1} \in \arg \min_{\bB } ~ g_k(\bB) + \lambda \, h(\bB \bY) ~ {\rm s.t.} ~ \bB^\top \bone =  (\bY^\top)^\dag \bone
\eeq 
for $k=0,1,2,\cdots$. Here,
\[
g_k(\bB) := f(\bB^k) + \langle \nabla f(\bB^k), \bB  - \bB^k \rangle + \mu_k \| \bB - \bB^k \|^2
\]
is a quadratic approximation of $f$ at $\bB^k$, where $\mu_k > 0$;
$\nabla f$ is the gradient of $f$;
$\| \cdot \|$ is the Euclidean norm.
Knowledge readers on optimization may notice that \eqref{eq:sca_sisal} resembles the proximal gradient method (see, e.g., \cite{beck2017first}),
although they should also be warned that $f$ does not have Lipschitz continuous gradient, a key assumption with the use of proximal gradient.
The question that remains is to solve the convex problems in \eqref{eq:sca_sisal}.
Bioucas-Dias devised a specialized algorithm for \eqref{eq:sca_sisal} via the variable splitting augmented Lagrangian method (which is the same as the alternating direction method of multipliers).
The algorithm exploits the problem structure and is computationally very efficient.

The iterations in \eqref{eq:sca_sisal} is the basic form of SISAL.
The actual algorithm is a modification that resembles the gradient projection method with the limited minimization step-size rule \cite[Ch.~2.3]{bertsekas1999nonlinear}.

SISAL shows Bioucas-Dias' strong application insights with what we call non-convex large-scale optimization today.
Bioucas-Dias did so in 2009, well ahead of the blooming of the topic in signal and image processing, machine learning, data science, etc.

\subsection{Some Well-Known Advantages of SISAL}

SISAL generally runs faster than other SVMin state-of-the-art schemes,
such as those that tackle the hard-constrained SVMin problems (such as \eqref{eq:svmin_jose2});
this is particularly so when $T$ is very large.
SISAL is known to be robust to noise and outliers.
In that regard we should first point out that the SVMin intuition, as well as the formulations \eqref{eq:svmin}--\eqref{eq:svmin_jose2}, were established on the case of noiseless data.
In the noisy case, 
using hard constraints to enforce enclosing of all the pixels can result in sensitivity issues, subsequently causing poor estimates of the endmembers.
Using soft constraints allows some pixels, particularly the adversarial ones, to lie outside the simplex,
and that tolerance can be beneficial in mitigating the sensitivity effects.
The only issue that is not easy to answer is the selection of $\lambda$ in \eqref{eq:svmin_jose3};
i.e., should we suppress volume more, or encourage non-negativity more?
Usually, the parameter $\lambda$ is manually chosen.

\subsection{Further Discussion}

In 2014, Guangzhou, Bioucas-Dias and the author discussed the SVMin identifiability which was solved in the noiseless case \cite{lin2015identifiability,Fu2015}.
Bioucas-Dias challenged the author with this difficult question.
\begin{Question}
	Ill-conditioned $\bA$ can cause serious noise sensitivity.
\end{Question}
This issue is valid in practice, and noise sensitivity analysis for SVMin is still an unsolved problem.
But recently the author has a different way to answer this question, which will be explained later.
In 2017, they discussed online the following problem.
\begin{Question}
Is minimizing $\svol(\bA) \propto (\det(\bar{\bA}^\top \bar{\bA}))^{1/2}$ over data-enclosing constraints, i.e., the SVMin problem \eqref{eq:svmin}, the same as the counterpart of minimizing $(\det(\bA^\top \bA))^{1/2}$, as in \eqref{eq:svmin_jose}?\footnote{The author favors $\svol(\bA)$ zealously because it is the true simplex volume. Bioucas-Dias prefers $(\det(\bA^\top \bA))^{1/2}$ because it is simpler.}
\end{Question}
While $\det(\bar{\bA}^\top \bar{\bA})$ and $\det(\bA^\top \bA)$ are similar, it does not mean that they lead to the same minimization result.
They studied this aspect and the answer is yes.
The proof is shown in Section~6.2.

\section{DECA}

DECA pursues a probabilistic paradigm for HU.
It is different from pure-pixel search and SVMin right from the onset.
But researchers may not know that DECA can be viewed as a more powerful form of SVMin, subsuming SVMin as a special case.
The author will describe DECA using his derivations \cite{PRISM2021}.
Consider a noisy model
\[
\by_t = \bA \bs_t + \bv_t,
\]
where $\bv_t$ is noise.
Assume the following:
i) $\bA$ has affinely independent columns;
ii) all the $\bs_t$'s and $\bv_t$'s are independent;
iii) every $\bv_t$ follows a Gaussian distribution with mean zero and covariance $\sigma^2 \bI$;
iv) every $\bs_t$ follows a Dirichlet mixture distribution, whose probability density function (PDF) is given by 
\[
p(\bs; \bm \gamma, \balp_1, \ldots, \balp_K ) = \textstyle \sum_{k=1}^K \gamma_k D(\bs;\balp_k)
\]
for some parameters $\bm \gamma > \bzero$, $\bm \gamma^\top \bone =1$, $\bm \alpha_1, \ldots, \bm \alpha_K > \bzero$; $D(\bs;\balp)$ denotes the Dirichlet distribution with concentration parameter $\balp$.
The use of the Dirichlet mixture prior  is to provide a general model for accommodating complex phenomena in real-world hyperspectral data.
The problem is to estimate $\bA$, together with the unknown prior parameters $\bm \gamma$ and $\balp_k$'s, by maximum-likelihood (ML) inference:
\beq \label{eq:ml}
\hat{\btheta}_{\sf ML} \in \arg \max_{\btheta \in \Theta} \textstyle \frac{1}{T} \sum_{t=1}^T \log p(\by_t;\btheta)
\eeq 
where $\btheta = \{ \bA, \bm \gamma, \balp_1, \ldots, \balp_K \}$;
$\Theta$ is the domain of $\btheta$;
$p(\by;\btheta)$ is the PDF of $\by$ parameterized by $\btheta$ and is given by
\begin{align}
p(\by;\btheta) & = \int p(\by| \bs; \bA) p(\bs; \bm \gamma, \balp_1, \ldots, \balp_K ) {\rm d}\mu(\bs) \nonumber \\
& = \textstyle \sum_{k=1}^K \gamma_k \int \varphi_\sigma(\by- \bA \bs) D(\bs;\balp_k) {\rm d}\mu(\bs) \label{eq:intract_int}
\end{align}
in which $\varphi_\sigma(\by) = e^{- \| \by \|^2/{2\sigma^2} }/( \sqrt{2 \pi} \sigma)^{M}$, and $\mu$ is the Lebesgue measure on $\{ \bs \in \Rbb^N \mid \bone^\top \bs = 1 \}$.
In statistical inference, ML estimation is known to have benign traits of properties, such as  better estimation accuracy for larger $T$, under some assumptions.

DECA considers the noiseless case, for which the integrals in \eqref{eq:intract_int} have closed form, and uses expectation maximization (EM) to realize the ML estimator \eqref{eq:ml}.
Instead of describing the EM algorithm, the author wants to draw connection.
Let us simplify by adopting
the uniform prior model, i.e., $p(\bs; \bm \gamma, \balp_1, \ldots, \balp_K ) = D(\bs; \bone)$.
Also, assume $M= N-1$. Then, as shown in \cite{PRISM2021}, the log likelihood for affinely independent $\bA$ can be expressed as
\beq
\log p(\by;\btheta) = -\log \svol(\bA) + \underbrace{\log\left(  \int_{\Rbb^{N-1}} \varphi_\sigma(\by - \bx )  \indfn{\bar{\setA}}(\bx) {\rm d}\bx  \right)}_{:= r(\bA, \by)},
\nonumber 
\eeq 
where 
$\indfn{\setX}(\bx) = 1$ if $\bx \in \setX$, $\indfn{\setX}(\bx) = 0$ if $\bx \notin \setX$;
$\setA = \conv(\bA)$; $\bar{\setA}$ is the relative interior of $\setA$.
The ML problem \eqref{eq:ml} then reduces to
\beq \label{eq:ml2}
\hat{\bA}_{\sf ML} \in \arg \min_{\bA} ~ \log \svol(\bA) - \textstyle \frac{1}{T} \sum_{t=1}^T r(\bA,\by_t).
\eeq 
Problem \eqref{eq:ml2} appears as a soft-constrained SVMin:
as $r(\bA,\by)$ is large (respectively, small) for $\by$ lying well inside (respectively, far away) the simplex $\setA$,
it serves a penalizer for outside-the-simplex points in \eqref{eq:ml2}.
In fact, for the noiseless case $\sigma^2 = 0$,
\beq \label{eq:logp_sig_0}
r(\bA,\by) = \left\{
\begin{array}{ll}
0, & \by \in \bar{\setA} \\
\infty, & \by \notin \bar{\setA}
\end{array} 
\right. 
\nonumber 
\eeq 
and the ML estimator \eqref{eq:ml2} becomes the SVMin problem \eqref{eq:svmin}!
This identity was informally mentioned by Bioucas-Dias in his WHISPERS 2009 presentation.\footnote{The author has been deeply intrigued by that since then.}
It was also alluded to in his paper \cite{nascimento2012hyperspectral},
though not apparent.
In this connection, it is worth recognizing that Dobigeon {\em et al.} described a similar result on the above-noted identity around the same time \cite[Appendix]{dobigeon2009joint}, although it is also not apparent.
The author's latest study \cite{PRISM2021} further reveals how SISAL can roughly be seen as an approximation of the ML estimator \eqref{eq:ml2} in the noisy case;
the details 
are omitted.
The author also shows that 
\begin{Theorem} \label{thm:id}
	Consider $T \rightarrow \infty$.
	The ML estimator \eqref{eq:ml2} can lead to exact identification of the true $\bA$.
	The result holds for the general noisy case.
\end{Theorem}
The above result is vital in confirming the strength of ML. 
It is worth noting that the DECA paper \cite{nascimento2012hyperspectral} provided an intuitive justification (but not a proof) on the same result.
Theorem~\ref{thm:id} also gives an impression that 
the noise effects should be reduced as $T$ increases.

The author now gives a partial answer to Question 1: 
If we have a large number of pixels (also known as big data),
we may mitigate the impact of noise by employing the ML estimator, or a good approximation of it by soft-constrained SVMin.



\section{Closing Remark}

The author wants to express his very heartfelt gratitude to Bioucas-Dias for his many inspirations, challenges and encouragements,
which led him to work on interesting problems.

\section{Appendix}

\subsection{Proof of Equivalence of Problems \eqref{eq:svmin_jose} and \eqref{eq:svmin_jose2}}

Let $(\bA,\bS)$ be any feasible point of \eqref{eq:svmin_jose}.
Redefine the true $(\bA,\bS)$ in the data model \eqref{eq:model} as $(\bA_0,\bS_0)$.
Since $\bA_0 \bS_0 = \bY = \bA \bS$,
and $\bA_0$ and $\bS_0^\top$ have full column rank,
one can show that $\bA$ and $\bS^\top$ have full column rank.
Consider $M= N$. 
Let $\bB = \bA^{-1}$.
We have 
$\bY = \bA \bS \Longleftrightarrow \bB \bY = \bS$.
Applying the above to $\bS^\top \bone = \bone$ yields
\beq \label{eq:subtle}
\bY^\top \bB^\top \bone = \bone ~ \Longrightarrow \bB^\top \bone = (\bY^\top)^\dag \bone.
\eeq 
One would be tempted to think that the converse of \eqref{eq:subtle} 
\beq \label{eq:subtle2}
\bY^\top \bB^\top \bone = \bone ~ \Longleftarrow \bB^\top \bone = (\bY^\top)^\dag \bone
\eeq 
is also true, but it is not true if we see the problem as a generic matrix analysis problem.
But \eqref{eq:subtle2} can be shown by incorporating $\bY= \bA_0 \bS_0$.\footnote{This is the part that is different from Bioucas-Dias' derivations.}
To put into context, consider finding an $\bx$ such that
\beq \label{eq:2subtle}
\bY^\top \bx  = \bone. 
\eeq 
If a solution to \eqref{eq:2subtle} exists, then $\bx$ is uniquely given by $\bx = (\bY^\top)^\dag \bone$.
Also, as a consequence, $\bx = (\bY^\top)^\dag \bone \Longrightarrow \bY^\top \bx  = \bone$  will be true.
One can verify that $\bx = \bA_0^{-\top} \bone$ satisfies \eqref{eq:2subtle} (we need $\bS_0^\top \bone = \bone$).
Thus, $\bx = (\bY^\top)^\dag \bone \Longrightarrow \bY^\top \bx  = \bone$ is true, and \eqref{eq:subtle2} is also true.

Using the above results, we can equivalently transform \eqref{eq:svmin_jose} to \eqref{eq:svmin_jose2}, with an extra constraint that $\bB$ is invertible.
We can discard the invertibility constraint from \eqref{eq:svmin_jose2} w.l.o.g., as $\det(\bB) = 0$ for non-invertible $\bB$.
Let us conclude:
{\em  Under the noiseless data model \eqref{eq:model} and the assumptions thereof,
problems \eqref{eq:svmin_jose} and \eqref{eq:svmin_jose2} are equivalent.}

\vspace{-.5em}
\subsection{Equivalence of Minimizing $\det(\bar{\bA}^\top \bar{\bA})$ and $\det(\bA^\top \bA)$}

We use the same convention as above:
$(\bA_0,\bS_0)$ is the ground truth,
and $(\bA,\bS)$ is any feasible point of \eqref{eq:svmin_jose}.
Let $\bF = [~ \bI ~ -\bone ~]^\top \in \Rbb^{N \times (N-1)}$,
$\bG = [~ \bF ~ \tfrac{1}{N} \bone ~] \in \Rbb^{N \times N}$.
It can be shown that $\det(\bG)= 1$.\footnote{Hint: 
By Gauss elimination one can show that $\bH = \bM_{N-1} \cdots \bM_1 \bG$, where $\bM_i = \bI + \be_N \be_i^\top$; 
$\bH = \bI + \sum_{j=1}^{N-1} \be_j \be_N^\top$; $\be_i$ is such that $[ \be_i ]_i = 1$, $[ \be_i ]_j = 0$ for $j \neq i$. 
Also, it holds that $\det(\bM_i) = 1$,  $\det(\bH) = 1$.
}
By letting $\bb = \frac{1}{N} \bA\bone$, and noting $\bar{\bA} = \bA \bF$,
\begin{subequations} 
\begin{align}
& \det(\bA^\top \bA)  = |\det(\bG)|^2 \det(\bA^\top \bA) = \det(\bG^\top \bA^\top \bA \bG)
\nonumber  \\
& ~~ = \det \left( \begin{bmatrix} \bar{\bA}^\top  \bar{\bA} & \bar{\bA}^\top \bb \\
\bb^\top \bar{\bA} & \| \bb \|^2 \end{bmatrix} \right)  \\
& ~~ = \det( \bar{\bA}^\top \bar{\bA}) \cdot ( \| \bb \|^2 - \bb^\top \bar{\bA} ( \bar{\bA}^\top  \bar{\bA} )^{-1} \bar{\bA}^\top \bb) \label{eq:tt_b} \\
& ~~ = \det( \bar{\bA}^\top \bar{\bA}) \cdot \min_{\bz} \| \bb - \bar{\bA} \bz \|^2. \label{eq:tt_c}
\end{align}
\end{subequations}
where \eqref{eq:tt_b} is due to Schur's determinant identity;
$\bar{\bA}^\top  \bar{\bA}$ is invertible because $\bA$ has full column rank (as shown in Section 6.1).
If $\min_{\bz} \| \bb - \bar{\bA} \bz \|^2$ is a constant $C > 0$ irrespective of $\bA$,
then we will have the desired result $\det(\bA^\top \bA) = C \det( \bar{\bA}^\top \bar{\bA})$.

To that end, denote $\aff(\bA) := \{ \by = \bA \bx \mid \bx^\top \bone = 1 \}$.
It can be verified that $\aff(\bA) = \aff(\bA_0)$  \cite{Chan2009}.
Let $\sspan( \bA )$ denote the span of the columns of $\bA$.
Consider the following.
\begin{Lemma} \label{lem:1}
	Let $\bA_0 \in \Rbb^{M \times N}$ be any matrix, let $\bar{\bA}_0 = \bA_0 \bF$,
	and let $\bU \in \Rbb^{N \times (N-1)}$ be a semi-orthogonal matrix such that $\bU^\top \bone = \bzero$.
	\begin{enumerate}[(a)]
		\item $\aff(\bA_0) = \sspan(\bar{\bA}_0) + \bd$ for any $\bd \in \aff(\bA_0)$
		\item $\sspan(\bar{\bA}_0)= \sspan(\bA_0 \bU)$
		\item $\{ \bx \in \Rbb^N \mid \bx^\top \bone = 1\} = \sspan(\bU) + \frac{1}{N} \bone$
	\end{enumerate}
\end{Lemma}
The proof of Lemma~\ref{lem:1} can be found in \cite{PRISM2021}.
Choose $\bd =  \tfrac{1}{N} \bA_0 \bone$.
From $\aff(\bA) = \aff(\bA_0)$,
we have $\sspan(\bar{\bA}) = \sspan(\bA_0 \bU)$.
Also, since $\bb \in \aff(\bA)$ we can write $\bb = \bA_0 \bU \tilde{\bz} + \bd$ for some $\tilde{\bz}$.
Then
\begin{align*}
& \min_{\bz} \| \bb - \bar{\bA} \bz \|^2 = \min_{\bz} \| \bb - \bA_0 \bU \bz \|^2 \\
& = \min_{\bz} \| \bA_0 ( \bU(\tilde{\bz} - \bz ) + {\textstyle \tfrac{1}{N} \bone}  ) \|^2
= \min_{\bx^\top \bone = 1 } \| \bA_0 \bx \|^2 > 0;
\end{align*}
(recall that $\bA_0$ has full column rank). The proof is done.
To summarize,
{\em the answer to Question~2 is yes under the noiseless data model \eqref{eq:model} and the assumptions thereof.}

{ 

\footnotesize
\bibliographystyle{IEEEtran}
\bibliography{refs}
}

\end{document}